\documentclass[twocolumn,prl]{revtex4}
\usepackage[a4paper, total={7in, 10in}]{geometry}
\usepackage[parfill]{parskip}
\usepackage{physics, tensor, float}
\usepackage{graphicx}
\graphicspath{ {Plots/} }
\usepackage{bm}
\usepackage{soul}
\usepackage{amssymb,amsmath,amsthm}
\usepackage{hyperref}
\usepackage{mathrsfs}
\usepackage[utf8]{inputenc}
\usepackage{enumerate}
\usepackage{bigints}
\usepackage{xcolor}
\usepackage{appendix}
\usepackage{graphicx}
\usepackage{float}
\usepackage{tikz}
\usepackage{setspace}
\usepackage{cancel}
\usepackage{doi}
\definecolor{purple}{rgb}{1,0,1}
\definecolor{lime}{HTML}{A6CE39} 

\definecolor{lime}{HTML}{A6CE39}
\newcommand{\orcidicon}{%
	\begin{tikzpicture}
	\draw[lime, fill=lime] (0,0) 
		circle [radius=0.16] 
		node[white] {{\fontfamily{qag}\selectfont \tiny ID}};
	\draw[white, fill=white] (-0.0625,0.095) 
		circle [radius=0.007];
	\end{tikzpicture}
	\hspace{-5mm}
}

\newcommand\orcidAlex{{\href{https://orcid.org/0000-0002-1763-3563}{\orcidicon}}}
\newcommand\orcidMatt{{\href{https://orcid.org/0000-0003-1088-6485}{\orcidicon}}}

\newcommand{\e}{\mathrm{e}}

\begin{document}

\title{
Astrophysically viable Kerr-like spacetime --- into the eye of the storm
}
\author{
Alex Simpson\!\orcidAlex\! and Matt Visser\!\orcidMatt\!
}
\affiliation{School of Mathematics and Statistics, Victoria University of Wellington,
\null\qquad PO Box 600, Wellington 6140, New Zealand.}
\email{alex.simpson@sms.vuw.ac.nz}
\email{matt.visser@sms.vuw.ac.nz}

\begin{abstract}

We analyse a rotating regular black hole with asymptotically Minkowski core. This Kerr-like geometry possesses the full ``Killing tower" of nontrivial Killing tensor, Killing-Yano tensor, and principal tensor. The Hamilton-Jacobi equation, the Klein-Gordon equation, and Maxwell's equations are separable. Energy-condition-violating physics is pushed into an arbitrarily small region in the deep core. The geometry has a very high level of mathematical tractability; extraction of astrophysical observables falsifiable/verifiable by the observational community is straightforward.

\bigskip
\noindent
{\sc Date:} 7 December 2021; \LaTeX-ed \today

axisymmetry, LIGO/Virgo, LISA, Killing tower, energy conditions.

%
\end{abstract}

\maketitle
\def\tr{{\mathrm{tr}}}
\def\diag{{\mathrm{diag}}}
\def\cof{{\mathrm{cof}}}
\def\pdet{{\mathrm{pdet}}}
\def\d{{\mathrm{d}}}
\parindent0pt
\parskip7pt
\def\Kerr{{\scriptscriptstyle{\mathrm{Kerr}}}}
\def\eos{{\scriptscriptstyle{\mathrm{eos}}}}
\emph{Introduction:} Observational technologies such as LIGO/\-Virgo~\cite{LIGO1}, and the Event Horizon Telescope~\cite{EHT1}, enable incredible new observational opportunities. The planned LISA project~\cite{LISA}, and planned next-generation ground-based observatories~\cite{science-book}, will have even more impressive scope. This gives impetus to the need to streamline the discourse between theory and experiment.

Since classical curvature singularities occur at distance scales where general relativity (GR) is no longer obviously applicable, one attractive option is to classically excise these singularities in astrophysically appropriate contexts, and extract quantities which are, in principle, observationally falsifiable/verifiable. This speaks directly to the ``traditional'' scientific method, and theoretically speaking one is simply modifying the axioms of GR by including a new one: ``forbid classical curvature singularities''. It is hoped that this will be a tractable path to progress in the absence of any  falsifiable/\-verifiable theory of quantum gravity. As our observational capacity increases, future measurements of astrophysical signatures will enable us to delineate between candidate spacetimes, giving exciting new insights.

Strategically, one should impose physics constraints which speak directly to the observational community. For instance, let us ask that an ``idealised" candidate spacetime satisfies the following:
\vspace{-5pt}
\begin{itemize}
\itemsep-6pt
    \item Impose axisymmetry; astrophysical sources rotate.
    \item Impose asymptotic flatness at spatial infinity~\cite{PGLT1}.
    \item Impose separability of the Hamilton--Jacobi (HJ) equations for ``in principle'' integrable geodesics able to be compared with observational data. (In axisymmetry a sufficient condition for this is the existence of a nontrivial Killing tensor $K_{\mu\nu}$).
    \item Impose separability of Maxwell's equations, and the equations governing spin two axial and polar modes. Invoke the inverse Cowling approximation, and analyze the permitted quasi-normal modes for spin one EM and spin two GR perturbations \emph{via} standard numerical techniques.
    \item Impose a high threshold of mathematical tractability. Mathematically tractable candidate spacetimes yield astrophysical observables with significantly more ease.
    \item Impose constraints on satisfaction/\-violation of the classical energy conditions; at the very least in the region outside the outer horizon. Empirically, violation of the energy conditions should only occur at quantum scales; apart from SEC-violations due to positive $\Lambda$, we have not observed exotic matter in an astrophysical context~\cite{Visser:epoch,Visser:epoch-prd,Visser:cosmo99}.
\end{itemize}
This list is a guideline and should be appropriately edited to reflect ongoing developments. Further desirable constraints could include forbidding closed timelike curves, or enforcing separability of the Dirac equation. However the list above speaks \emph{directly} to currently available observations. Developing geometries which manifestly satisfy these constraints is highly nontrivial, and should be an area of focus for the GR community.

One subset of curvature-singularity-free geometries are the ``regular black holes" (RBHs). We mean ``regular'' in the sense of Bardeen~\cite{Bardeen:1968}, achieved by enforcing global finiteness on Riemann curvature invariants and orthonormal curvature tensor components. Herein we analyse a rotating RBH with asymptotically Minkowski core. The geometry was apparently first proposed by Ghosh~\cite{Ghosh:2014}; we developed it independently (and analyzed it in much more detail) in reference~\cite{Eyeofstorm}. This spacetime is a tightly controlled deviation from Kerr, still set in stationary axisymmetry, and is asymptotically Kerr for large $r$. It has a \emph{very} high level of mathematical tractability, possessing the full ``Killing tower''~\cite{Frolov:2017} of nontrivial Killing tensor $K_{\mu\nu}$, Killing--Yano tensor, and principal tensor. The Killing tensor induces a Carter constant~\cite{Carter:1968a,Carter:1968b}, giving in principle integrable geodesics (\emph{i.e.}, HJ-separability). Any energy-condition-violating physics can be pushed  into the deep core, at a distance scale where GR is no longer an appropriate theory. Both the Klein--Gordon and Maxwell's equations separate.

With respect to the above list of proposed constraints, this geometry is as close to ``observationally ideal'' as we have found in the literature.

\emph{Metric ansatz:} The candidate spacetime analysed herein is inspired directly by the regular black hole with asymptotically Minkowski core analysed in reference~\cite{Simpson:2020}. That line element is given explicitly by:
\begin{equation}\label{AM}
    \d s^2 = -\left(1-\frac{2m\,\e^{-\ell/r}}{r}\right)\,\d t^2 + \frac{\d r^2}{1-\frac{2m\,\e^{-\ell/r}}{r}} + r^2\,\d\Omega^2_2 \ .
\end{equation}
Schwarzschild spacetime, in standard curvature coordinates, is recovered precisely in the limit as $\ell\rightarrow0$. The parameter $\ell$ can be viewed as quantifying the deviation from Schwarzschild spacetime. This  ``regularized'' black hole can be obtained from  Schwarzschild by simply replacing:
\begin{itemize}
    \item $m\rightarrow m(r)=m\,\e^{-\ell/r}$.
\end{itemize}
Note that this is \emph{not} a coordinate transformation.
The region $r<0$ is unphysical for this candidate spacetime due to the severe mathematical discontinuity of $\e^{-\ell/r}$ at $r=0$. Physical analysis is valid for $r\geq0$.

We wish to investigate a rotating version of this regular black hole with asymptotically Minkowski core. Begin with Kerr in Boyer--Lindquist (BL) coordinates:
\begin{eqnarray}\label{Kerr}
    \d s^2 &=& - \frac{\Delta_{\Kerr}}{\Sigma}(\d t-a\sin^2\theta\,\d\phi)^2
    +\frac{\Sigma}{\Delta_{\Kerr}}\,\d r^2 + \Sigma\,\d\theta^2
     \nonumber \\
     && \qquad \qquad \qquad + \frac{\sin^2\theta}{\Sigma}[(r^2+a^2)\,\d\phi-a\,\d t]^2 \ ,
\end{eqnarray}
where
\begin{equation}
    \Sigma = r^2+a^2\cos^2\theta \ , \qquad \Delta_{\Kerr}=r^2+a^2-2mr \ .
\end{equation}
%
%
In BL coordinates, Kerr possesses a ring singularity at $r=0$, which we wish to excise. Inspired by the aforementioned ``regularisation" procedure, we make the modification $m\rightarrow m(r)=m\,\e^{-\ell/r}$. Mathematically there is still a discontinuity at $r=0$, and hence $r<0$ is omitted from the analysis. In Cartesian coordinates we have $r_{naive}^2 = x^2+y^2+z^2$, and
\begin{equation}
r_{naive}^2 = r^2 + a^2 - \frac{a^2 z^2}{r^2} \ ; \qquad \cos\theta= z/r \ .
\end{equation}
Then
\begin{equation}
r_{naive}^2 = r^2 + a^2 \sin^2\theta \ ,
\end{equation}
while
\begin{eqnarray}
\cos\theta_{naive} &=& \frac{z}{r_{naive}} = \frac{z}{r} \;\frac{r}{r_{naive}} \nonumber \\
&=& 
\frac{\cos\theta}{\sqrt{ 1 + a^2 \sin^2\theta/r^2}}
=
\frac{r\; \cos\theta}{\sqrt{ r^2 + a^2 \sin^2\theta}} \,.\qquad
\end{eqnarray}
As $r\to 0^+$, exponential suppression in the BL  coordinate $r$ suppresses the mass function $m(r)$ in the entire Cartesian disk $r_{naive} \leq a$, where $\cos\theta_{naive}=0$. The closed timelike curves associated with the ``other universes'' in the maximal analytic extension of standard Kerr are removed from the analysis due to the restriction $r\geq0$.

\emph{The eye of the storm:} Accordingly, we advocate for the following fully explicit metric, dubbed the ``eye of the storm'' (eos) spacetime:
\begin{eqnarray}\label{E:eos}
    \d s^2 &=& \frac{\Sigma}{\Delta_{\eos}}\,\d r^2 + \Sigma\,\d\theta^2 - \frac{\Delta_{\eos}}{\Sigma}(\d t-a\sin^2\theta\,\d\phi)^2 \nonumber \\
    && \quad \qquad \qquad + \frac{\sin^2\theta}{\Sigma}[(r^2+a^2)\,\d\phi-a\,\d t]^2 \ ,
\end{eqnarray}
where
\begin{equation}
    \Sigma = r^2+a^2\cos^2\theta \ , \qquad \Delta_{\eos}=r^2+a^2-2mr\,\e^{-\ell/r} \ .
\end{equation}
%
%
Asymptotic flatness as $r\rightarrow+\infty$ is preserved. One recovers standard Kerr in BL coordinates as $\ell\rightarrow0$. We hence enforce $\ell>0$ for nontrivial analysis, and $\ell$ can be viewed as quantifying the deviation from Kerr. One recovers equation~(\ref{AM}) precisely in the limit as $a\rightarrow0$. Most importantly, the ring singularity is excised, replaced by an asymptotically Minkowski region of spacetime. As we shall shortly prove, this renders the geometry globally nonsingular --- we have a tractable model for a rotating regular black hole.

Ordering the coordinates as $(t,r,\theta,\phi)$, from equation~(\ref{E:eos}) it is trivial to read off a convenient covariant tetrad (co-tetrad) which is a solution of $g_{\mu\nu}=\eta_{\hat{\mu}\hat{\nu}}\;e^{\hat{\mu}}{}_{\mu}\; e^{\hat{\nu}}{}_{\nu}$ (note that this co-tetrad is not unique):
\begin{eqnarray}
    e^{\hat{t}}{}_{\mu} &=& \sqrt{\frac{\Delta_{\text{eos}}}{\Sigma}}(-1;0,0,a\sin^2\theta);\;\;
    e^{\hat{\theta}}{}_{\mu}= \sqrt{\Sigma}(0;0,1,0); 
    \nonumber \\
    e^{\hat{r}}{}_{\mu} &=& \sqrt{\frac{\Sigma}{\Delta_{\text{eos}}}}\left(0;1,0,0\right);\;
    e^{\hat{\phi}}{}_{\mu} = \frac{\sin\theta}{\sqrt{\Sigma}}
    (-a;0,0,r^2+a^2). \nonumber \\
    &&
\end{eqnarray}
The (contravariant) tetrad is then uniquely defined \emph{via} $e_{\hat{\mu}}{}^{\mu}=\eta_{\hat{\mu}\hat{\nu}}\; e^{\hat{\nu}}{}_{\nu}\; g^{\nu\mu}$. Explicitly:
\begin{eqnarray}
    e_{\hat{t}}{}^{\mu} &=& -\frac{1}{\sqrt{\Sigma\Delta_{\text{eos}}}}(r^2+a^2;0,0,a);\,
    e_{\hat{\theta}}{}^{\mu} = \frac{1}{\sqrt{\Sigma}}(0;0,1,0); 
    \nonumber \\
    e_{\hat{r}}{}^{\mu} &=& \sqrt{\frac{\Delta_{\text{eos}}}{\Sigma}}  
    (0;1,0,0); \;
    e_{\hat{\phi}}{}^{\mu} = \frac{1}{\sqrt{\Sigma}\sin\theta}
    (a\sin^2\theta;0,0,1). \nonumber \\
    &&
\end{eqnarray}
This tetrad will be used to convert tensor coordinate components into an orthonormal basis.

Let us analyse the nonzero components of the Riemann curvature tensor with respect to this orthonormal basis, to confirm that the eye of the storm is regular in the sense of Bardeen.
%
%
All such components are of the form
\begin{equation}
    R^{\hat{\alpha}\hat{\beta}}{}_{\hat{\mu}\hat{\nu}} = \frac{m\, \e^{-\ell/r}}{r^n\,\Sigma^3}X(r,\theta; a, \ell) \ ,
\end{equation}
where $X(r,\theta;a,\ell)$ is globally finite. 

The only potentially concerning behaviour comes from the $r^n\,\Sigma^3$ in the denominators, in the limit as $r\rightarrow0^{+}$. However $\lim_{r\rightarrow0^{+}}\e^{-\ell/r}/(r^n\,\Sigma^3)=0$ for all $\theta$. The ring singularity present at $r=0$ for Kerr in BL coordinates is replaced by an asymptotically Minkowski region of spacetime. This is enough to conclude that the spacetime is globally regular, as asserted.

\emph{Geometric analysis:} We define the useful object
\begin{equation}
    \Xi = \frac{\ell\Sigma}{2r^3} \ .
\end{equation}
%
Note that $\Xi$ is dimensionless.\\
 Curvature quantities will be displayed in the form:\\
 \leftline{$\hbox{(something dimensionful)} \times \hbox{(something dimensionless)}$.}

\emph{Curvature invariants:} Let us examine the behaviour of the Riemann curvature invariants. Ricci scalar:
\begin{equation}
    R = \frac{2\ell^2m\,\e^{-\ell/r}}{\Sigma\,r^3} \ .
\end{equation}
Ricci contraction $R_{\alpha\beta}R^{\alpha\beta}$:
\begin{equation}
    R_{\alpha\beta}R^{\alpha\beta} = \frac{8\ell^2m^2\,\e^{-2\ell/r}}{\Sigma^4}\left(\Xi^2-2\Xi+2\right) \ .
\end{equation}
Note positivity: $\left(\Xi^2-2\Xi+2\right) = 1 + (\Xi-1)^2 \geq 1$.

Kretschmann scalar ($K=R_{\alpha\beta\mu\nu}R^{\alpha\beta\mu\nu}$):
\begin{eqnarray}
    K &=& \frac{48r^6m^2\,\e^{-2\ell/r}}{\Sigma^6}\Bigg\lbrace 1 -15\frac{a^2}{r^2}\cos^2\theta +15\frac{a^4}{r^4}\cos^4\theta \nonumber \\
    && -\frac{a^6}{r^6}\cos^6\theta
    +\frac{4}{3}\Xi^4-\frac{16}{3}\Xi^3+8\Xi^2\left[1-\frac{a^2}{r^2}\cos^2\theta\right] \nonumber \\
    && \qquad \qquad -4\Xi\left[1-6\frac{a^2}{r^2}\cos^2\theta+\frac{a^4}{r^4}\cos^4\theta\right]\Bigg\rbrace.\quad
\end{eqnarray}
%
%
The orthonormal Weyl tensor has only two algebraically independent components
\begin{eqnarray}
C_{\hat t\hat\phi\hat t \hat\phi} &=& -\frac{C_{\hat t\hat r\hat t \hat r}}{2} =
C_{\hat t\hat\theta \hat t \hat\theta} = -C_{\hat\phi\hat r\hat \phi\hat r} 
= \frac{C_{\hat\phi\hat \theta\hat \phi\hat \theta}}{2} 
= -C_{\hat r \hat \theta\hat r \hat \theta}\,; \nonumber \\
&& \\
C_{\hat t\hat \phi\hat r\hat\theta} &=& \frac{C_{\hat t\hat r\hat\phi\hat\theta}}{2}
 = C_{\hat t\hat \theta\hat\phi\hat r} \,,
\end{eqnarray}
where explicitly
\begin{eqnarray}
C_{\hat t\hat\phi\hat t \hat\phi} &=& \frac{r^3m\,\e^{-\ell/r}}{3\Sigma^3}\left\lbrace 2\Xi^2-6\Xi+3-9\frac{a^2}{r^2}\cos^2\theta\right\rbrace \ ; \nonumber \\
&& \\
C_{\hat t\hat r\hat\phi\hat\theta} &=& \frac{r^2m\,\e^{-\ell/r}a\cos\theta}{\Sigma^3}\left\lbrace2\Xi-3+\frac{a^2}{r^2}\cos^2\theta\right\rbrace \,.
\end{eqnarray}
The quadratic Weyl contraction 
is then:
\begin{equation}
    C_{\alpha\beta\mu\nu}C^{\alpha\beta\mu\nu} = 
    48 ([C_{\hat t\hat\phi\hat t \hat\phi}]^2 
    - [C_{\hat t\hat \phi\hat r\hat\theta}]^2) \ .
\end{equation}
%
%
%
All invariants are globally finite. The standard results for Kerr spacetime are recovered in the limit as $\ell\rightarrow0$. All invariants tend to their counterparts for the spherically symmetric candidate geometry analysed in reference~\cite{Simpson:2020} in the limit as $a\rightarrow0$.
%

\emph{Ricci and Einstein tensors:} These  tensors diagonalize in the orthonormal basis. The Ricci tensor is
\begin{equation}
    R^{\hat{\mu}}{}_{\hat{\nu}} = \frac{2\ell m\,\e^{-\ell/r}}{\Sigma^2}\,\text{diag}\left(\Xi-1,\Xi-1,1,1\right) \ ,
\end{equation}
and the Einstein tensor is
\begin{eqnarray}\label{Einstein}
    G^{\hat{\mu}}{}_{\hat{\nu}} = -\frac{2\ell m\,\e^{-\ell/r}}{\Sigma^2}\,\text{diag}\left(1,1,\Xi-1,\Xi-1\right) \ .
\end{eqnarray}
When compared with other rotating regular black holes, these results are \emph{highly} tractable.

\emph{Causal structure and ergoregion:} The only coordinate singularities in the line element equation~(\ref{E:eos}) are located at the roots of $\Delta_{\eos}(r)$, which also characterise the horizon locations. There are either zero roots, one double root, or two distinct roots. $\Delta_{\eos}(r) > \Delta_{\Kerr}(r)$ implies that the root locations are trivially bounded by the locations of the roots of $\Delta_{\Kerr}(r)$. Specifically
\begin{equation}
m-\sqrt{m^2-a^2} < r_H^- \leq r_H^+ < m+\sqrt{m^2-a^2} \ .
\end{equation}
In particular, there are no roots if $m<a$.

We cannot analytically solve for the roots of $\Delta_{\eos}(r)$. However, assuming the existence of distinct roots $r_H^\pm$,  we can ``reverse engineer'' by solving for 
$m(r_H^+,r_H^-)$ and $a^2(r_H^+,r_H^-)$. By definition
\begin{equation}
   (r_H^+)^2  -2 m \, r_H^+ \exp( - \ell/r_H^+) + a^2 =0 \ ;
\end{equation}
\begin{equation}
   (r_H^-)^2  -2 m \, r_H^- \exp( - \ell/r_H^-) + a^2 =0 \ ,
\end{equation}
giving two simultaneous equations linear in $a^2$ and $m$. We find
\begin{equation}
m(r_H^+,r_H^-) =  \frac{(r_H^+)^2-(r_H^-)^2}{2 ( \e^{-\ell/r_H^+} \; r_H^+ - \e^{-\ell/r_H^-} \; r_H^-)} \ ,
\end{equation}
and
\begin{equation}
a^2 (r_H^+,r_H^-) = \frac{r_H^+ r_H^-  (
\e^{-\ell/r_H^-} \; r_H^+ - \e^{-\ell/r_H^+} \; r_H^-)}{\e^{-\ell/r_H^+} \; r_H^+ - \e^{-\ell/r_H^-} \; r_H^-} \ .
\end{equation}
The degenerate extremal limit $r_H^+\to r_H \leftarrow r_H^-$ simplifies \emph{via} l'H\^opital to
\begin{equation}
m(r_H) =  \frac{r_H^2 \; e^{\ell/r_H}}{r_H + \ell} > r_H  ; \;
a^2 (r_H) = \frac{r_H^2  (r_H-\ell)}{r_H+\ell} < r_H^2  .
\end{equation}
Setting $a\to a(r_H)$, for fixed $\ell$ and $r_H$ we have: (1) For $m>m(r_H)$ there are two distinct roots, one below and one above $r_H$. (2) For $m=m(r_H)$ there is one degenerate root at $r_H$. (3) For $m<m(r_H)$ there are no real roots.

%
Given that in this context the parameter $\ell$ is typically associated with the Planck scale, it is useful to perform a Taylor series expansion about $\ell=0$. Let us write
\begin{equation}
    r_{H} = m\,\e^{-\ell/r_{H}^\pm} + S_1 \sqrt{m^2\,\e^{-2\ell/r_{H}}-a^2} \ ,
\end{equation}
where $S_1=\pm 1$ depending on whether one is at an outer/inner horizon respectively. To second-order for small $\ell$ we find
\begin{equation}
    r_{H} = m + S_{1}\sqrt{m^2-a^2-2m\ell-\mathcal{O}(\ell^2)} \ .
\end{equation}
This has the correct behaviour as $\ell\rightarrow0$. 
%
%
%
%

$g_{tt}=0$ characterises the ergosurface, given \emph{implicitly} by
\begin{equation}
    r_{\text{erg}}^2+a^2\cos^2\theta = 2mr_{\text{erg}}\,\e^{-\ell/r_{\text{erg}}} \ .
\end{equation}
To second-order for small $\ell$
\begin{equation}
    r_{\text{erg}} = m + \sqrt{m^2-a^2\cos^2\theta-2m\ell-\mathcal{O}(\ell^2)} \ .
\end{equation}
This has the correct behaviour as $\ell\rightarrow0$. 
%
%
%
%

\emph{Killing tensor and Killing tower:}\label{Killingtower}
%
%
%
In the orthonormal tetrad basis, one can define the object
\begin{equation}
    K^{\hat{\mu}\hat{\nu}} = \begin{bmatrix}-a^2\cos^2\theta & 0 & 0 & 0\\ 0 & a^2\cos^2\theta & 0 & 0\\ 0 & 0 & -r^2 & 0\\ 0 & 0 & 0 & -r^2\\\end{bmatrix}^{\hat{\mu}\hat{\nu}} \ .
\end{equation}
Lowering the indices, it is straightforward to verify (\emph{e.g.}, using {\sf Maple}) that this results in a nontrivial Killing $2$-tensor satisfying $K_{({\hat{\mu}}{\hat{\nu}};{\hat{\alpha}})}=0$. Notice that in the tetrad basis this is identical to the nontrivial Killing 2-tensor for standard Kerr spacetime.

Furthermore, it is straightforward (\emph{e.g.}, using {\sf Maple}) to verify that the following two-form square-root of the Killing tensor is a genuine Killing--Yano tensor, satisfying the Killing--Yano equation $f_{\hat{\mu}(\hat{\nu};\hat{\alpha})}=0$:
\begin{equation}
    f^{\hat{\mu}\hat{\nu}} = \begin{bmatrix} 0 & a\cos\theta & 0 & 0\\ -a\cos\theta & 0 & 0 & 0\\ 0 & 0 & 0 & -r\\ 0 & 0 & r & 0\\\end{bmatrix}^{\hat{\mu}\hat{\nu}} \ .
\end{equation}
It is easily checked that $K^{\hat{\mu}\hat{\nu}} = -f^{\hat{\mu}\hat{\alpha}}\;\eta_{\hat{\alpha}\hat{\beta}}\;f^{\hat{\beta}\hat{\nu}}$. The ``principal tensor''~\cite{Frolov:2017} is then simply the Hodge dual of this two-form.

Specifically, notice that all three objects in the full ``Killing tower" are \emph{independent} of the  mass function $m(r)$. Hence, if one defines the family of ``$1$-function off-shell'' Kerr geometries by modifying standard BL Kerr \emph{via} $m\rightarrow m(r)$, all of these geometries inherit the full ``Killing tower" from Kerr.

Notably, both the Ricci tensor and Killing tensor are diagonal in this tetrad basis, and as such the commutator $[R,K]^{\hat{\mu}}{}_{\hat{\nu}}$ will vanish; 
this is sufficient to conclude that the Klein--Gordon equation is separable on the background spacetime~\cite{PGLT2}. Furthermore, we can conclude from the proof in Appendix A of reference~\cite{Franzin:2021a} that given this eye of the storm geometry falls within the ``$1$-function off-shell" Kerr class, Maxwell's equations will also be separable on the background spacetime. As such, the eye of the storm is amenable to both a standard spin zero and spin one quasi-normal modes analysis (invoke the inverse Cowling approximation, assume a separable wave form, and use your favourite numerical technique to approximate the ringdown signal).

Possession of the full ``Killing tower" as well as separability of both the Klein--Gordon and Maxwell equations are highly desirable features of this eye of the storm geometry. We conjecture that the equations governing the spin two polar and axial modes will also separate on the geometry; for now this is a topic for future research.

%
\emph{Stress-energy and energy conditions:}
%
%
We interpret the spacetime through the lens of standard GR, fixing the geometrodynamics. Coupling the geometry to the Einstein equations yields
\begin{equation}
    \frac{1}{8\pi}G^{\hat{\mu}}{}_{\hat{\nu}} = T^{\hat{\mu}}{}_{\hat{\nu}} = \text{diag}(-\rho,p_{r},p_{t},p_{t}) \ .
\end{equation}
Note $-\rho=p_{r}$ 
holds globally, 
regardless of whether one is outside (inside) the outer (inner) horizon, or trapped in between them. The diagonal orthonormal Einstein tensor implies that the stress-energy tensor is Hawking--Ellis type I~\cite{Hawking:1973}. From equation~(\ref{Einstein}), this gives the following stress-energy components:
\begin{eqnarray}
    \rho &=& -p_{r} = \frac{\ell m\,\e^{-\ell/r}}{4\pi\Sigma^2} \ ; \nonumber \\
    p_{t} &=& \frac{\ell m\,\e^{-\ell/r}}{4\pi\Sigma^2}\left(1-\Xi\right) \ .
\end{eqnarray}
Let us examine the classical energy conditions associated with GR. $\ell>0$ means that we trivially satisfy 
$\rho>0$ globally. $\rho+p_{r}=0$ globally, implying the radial null energy condition (NEC) is trivially satisfied. Analysing the transverse NEC:
\begin{equation}
    \rho + p_{t} = \frac{\ell m\,\e^{-\ell/r}}{4\pi\Sigma^2}\left(2-\Xi\right) \ .
\end{equation}
This changes sign when $\Xi=2$, \emph{i.e.} $\frac{\Sigma}{r^3}=\frac{4}{\ell}$. For $\theta=\pi/2$ this is when $r=\frac{\ell}{4}$. If $\frac{\Sigma}{r^3}<\frac{4}{\ell}$, the transverse NEC is satisfied, whilst if $\frac{\Sigma}{r^3}>\frac{4}{\ell}$, it is violated. For $\theta=\pi/2$, the violation region is when $r<\frac{\ell}{4}$.

The strong energy condition (SEC) involves:
\begin{equation}
    \rho + p_{r} + 2p_{t} = 2p_{t} = \frac{\ell m\,\e^{-\ell/r}}{2\pi\Sigma^2}\left(1-\Xi\right) \ .
\end{equation}
This changes sign when $\Xi=1$, \emph{i.e.} $\frac{\Sigma}{r^3}=\frac{2}{\ell}$. If $\frac{\Sigma}{r^3}<\frac{2}{\ell}$, the SEC is satisfied, whilst if $\frac{\Sigma}{r^3}>\frac{2}{\ell}$, the SEC is violated. For $\theta=\pi/2$ the violation region is whenever $r<\frac{\ell}{2}$.

Given the freedom to scale $\ell$, all of the energy-condition-violating physics can be pushed into an arbitrarily small region in the deep core,
%
%
where GR is no longer an appropriate theory. No exotic matter is required in the domain of outer communication. Outside horizons, we manifestly satisfy all of the classical energy conditions. This is consistent with astrophysical observations. When compared with the literature on rotating RBHs, the relationship with the classical energy conditions is an extremely desirable feature of eye of the storm spacetime; often rotating RBHs \emph{globally} violate several energy conditions.

\emph{Conclusions:}\label{S:discussion} The eye of the storm spacetime was constructed according to a set of carefully chosen theoretically and observationally motivated criteria. Analysing the geometry, it models a rotating regular black hole with asymptotically Minkowski core, is asymptotically Kerr at large distances, satisfies all of the standard energy conditions of GR in the regions of theoretical and observational validity, has integrable geodesics, and has the property of separability of both the Klein--Gordon and Maxwell's equations. It possesses the full ``Killing tower" of principal tensor, Killing--Yano tensor, and nontrivial Killing tensor. The eye of the storm is seemingly the most mathematically tractable rotating RBH in the literature, and is readily amenable to the extraction of astrophysical observables in principle falsifiable/\-verifiable by the observational community.

Separability of Klein--Gordon and Maxwell's equations indicates one can and should probe the geometry \emph{via} quasi-normal mode analysis for spin zero and spin one perturbations. Testing the same for the equations governing spin two axial and polar modes is an important future research topic. One should also look to extract the geodesics for test particles in the spacetime.

%
\emph{Acknowledgements:} AS was supported by a Victoria University of Wellington PhD scholarship, and was also indirectly supported by the Marsden Fund, \emph{via} a grant administered by the Royal Society of New Zealand. AS would also like to thank Ratu Mataira for useful conversations and discussion. MV was directly supported by the Marsden Fund, \emph{via} a grant administered by the Royal Society of New Zealand.


\end{document}